\documentclass{aa}

\usepackage{amsmath}
\usepackage{float}
\usepackage{txfonts}
\usepackage{hyperref}

\usepackage[usenames,dvipsnames,svgnames,table]{xcolor}
\definecolor{electricgreen}{rgb}{0.0, 1.0, 0.0}

\usepackage{appendix}
\usepackage[english]{babel}
\usepackage{comment}
\usepackage{graphicx}
\usepackage{natbib}
\usepackage{subfig}
\usepackage{multirow}
\usepackage{booktabs}

\begin{document}

\title{VUV spectroscopy of carbon dust analogs: \\ contribution to interstellar extinction}

\author{L. Gavilan\inst{1} 
          \and
          I. Alata\inst{1,2}
          \and
          K. C. Le \inst{2}
         \and
          T. Pino\inst{2}
         \and
          A. Giuliani\inst{3}
        \and
          E. Dartois\inst{1}}

\institute{Institut d\textsc{\char13}Astrophysique Spatiale (IAS), CNRS,  Univ. Paris Sud, Universit\'e Paris-Saclay, F-91405 Orsay, France
         \and
         Institut des Sciences Mol\'eculaires d'Orsay (ISMO), CNRS, Univ. Paris Sud, Universit\'e Paris-Saclay, F-91405 Orsay, France
           \and
          DISCO beam line, SOLEIL synchrotron, Saint Aubin, France \\
             \email{lisseth.gavilan@ias.u-psud.fr}
           }

\abstract {A full spectral characterization of carbonaceous dust analogs is necessary to understand their potential as carriers of observed astronomical spectral signatures such as the ubiquitous UV bump at 217.5 nm and the far-ultraviolet (FUV) rise common to interstellar extinction curves.} 
{Our goal is to study the spectral properties of carbonaceous dust analogs from the FUV to the mid-infrared (MIR) domain. We seek in particular to understand the spectra of these materials in the FUV range, for which laboratory studies are scarce.} 
{We produced analogs to carbonaceous interstellar dust encountered in various phases of the interstellar medium: amorphous hydrogenated carbons (a-C:H), for carbonaceous dust observed in the diffuse interstellar medium, and soot particles, for the polyaromatic component. Analogs to a-C:H dust were produced using a radio-frequency plasma reactor at low pressures, and soot nanoparticles films were produced in an ethylene (C$_2$H$_4$) flame. We measured transmission spectra of these thin films (thickness $<$ 100 nm) in the far-ultraviolet (190 - 250 nm) and in the vacuum-ultraviolet (VUV; 50 - 190 nm) regions using the APEX chamber at the DISCO beam line of the SOLEIL synchrotron radiation facility. These were also characterized through infrared microscopy at the SMIS beam line.}  
 {We successfully measured the transmission spectra of these analogs from $\lambda$ = 1 $\mu$m to 50 nm. From these, we extracted the laboratory optical constants via Kramers-Kronig inversion. We used these constants for comparison to existing interstellar extinction curves.} {We extend the spectral measurements of these types of carbonaceous analogs into the VUV and link the spectral features in this range to the 3.4 $\mu$m band. We suggest that these two materials might contribute
to different classes of interstellar extinction curves.}


\maketitle

\section{Introduction}

\subsection{General overview}

Carbon-based materials are abundant in circumstellar environments and in the diffuse interstellar medium (DISM) \citep{vandiedenhoven2004,Henning1998, Greenberg1999,Draine2003,Jager2011}. They are mainly produced in the envelopes of asymptotic giant branch (AGB) stars \citep{Gail2009, Molster2003} and are injected into the DISM.  
Carbonaceous dust can then be processed (by photons, ions, cosmic rays) and can further evolve during their lifetime in these heterogeneous environments \citep{Jones2014a}. Carbon materials can be classified according to their sizes, which range from (large) polycyclic hydrocarbon molecules to solid-state carbon grains, their aliphatic vs aromatic content, and their spectra, among other properties (e.g., \cite{Henning1998}). In solid carbon, sp$^2$, sp$^3$, and mixed hybridizations are possible, giving rise to materials with different short-, medium-, and long-range orders \citep{Jager1998, Papoular1996, Tielens2008}. 

Different carbon dust components in space are identified by their observed spectral signatures. In the infrared, observations of the 3.4 $\mu$m absorption band were attributed to vibrational CH$_2$ and CH$_3$ aliphatic modes in carbonaceous materials. The counterparts of the 3.4 $\mu$m band at 6.85 $\mu$m and 7.25 $\mu$m were later observed by space telescopes. The carriers of these bands were named HAC by the astrophysical community \citep[e.g.,][]{Pendleton2002, Furton1999, Scott1997, Duley1994}. The HAC denomination for laboratory analogs has been used for a wide variety of hydrocarbon solids, even though the structure of these materials varies from amorphous to disordered carbons. \\
\indent In the following, we refer to the plasma-produced analogs as a-C:H, a term used in the physics community. Amorphous refers to the lack of order at any scale in these materials. The laboratory infrared spectra of these materials match observed spectra, particularly regarding the shape and relative strength of the 3.4 $\mu$m band \citep{Dartois2007}. The structure associated with the hydrogen-rich a-C:H materials is characterized by few aromatic units linked in a mainly aliphatic backbone (e.g., Fig. 5 of \cite{Dartois2005}). 

Some astrophysical carbon materials subjected to UV light re-emit in the mid-infrared. Observations of these emission features, called AIB for aromatic infrared bands, appear consistently between 3 to 20 $\mu$m and have been classified into three main classes \citep{Peeters2002}. The intimate structures of the carriers of these different classes of AIBs are still unknown, although they are probably built on a polyaromatic skeleton spanning sizes from large molecules to small nanograins \citep{Leger1984, Allamandola1985,  Goto2007, Sloan2007, Pino2008, Tielens2008, Joblin2011, Li2012, Carpentier2012, Yang2013}. 

Interstellar extinction curves, derived from observations of several lines of sight from the near-IR to the UV, provide another clue to the type of dust present in the ISM. A persistent spectral feature is the strong absorption bump at 4.6 $\mu$m$^{-1}$ (217.5 nm), discovered by \cite{Stecher1965}, whose central position is surprisingly constant over several lines of sight, and whose origin is still debated \citep{Fitzpatrick2005, Fitzpatrick2007}. A steep far-ultraviolet (FUV) rise also characterizes many of these curves, tracing extinction variations. The lack of correlation between the spectral properties of the UV bump and the relative strengths of the bump and the FUV rise points to different types of dust carriers \citep{Greenberg1983, Fitzpatrick2007}. Graphite particles were initially proposed as the carrier of the UV bump by \cite{Draine1984}. More recently, hydrogenated carbons were proposed as potential carriers \citep{Colangeli1997, Schnaiter1998, Mennella1999, Jager2008}, supporting theoretical work by \cite{Duley1998}.

The central goal of this study is to obtain \textit{\textup{absolute}} spectral properties of carbon dust analogs in both the VUV/Vis regions and the infrared ranges, complementing earlier studies of hydrocarbon solids, particularly in the VUV range, where optical measurements are scarce. A full spectral characterization of these materials will clarify their potential contributions as carriers of observed spectral features in the infrared and in the UV, such as the FUV rise and also the 217.5 nm bump of interstellar extinction curves. In addition, infrared spectra allow us to characterize the quality and suitability of our samples as interstellar analogs.  These transmission spectra allow the derivation of optical constants from these carbonaceous materials, necessary to simulate observed astronomical extinction curves. 

Carbon analogs have been produced in the laboratory using different methods and their spectra in transmission, reflection, or electron energy loss spectroscopy (EELS) (see \cite{Henning2004} for a review) has been used to extract optical constants. 
\cite{Stecher1965} first derived optical constants of nanometer-sized graphitic particles.
\cite{Draine1984} derived the optical constants from laboratory and observational data for graphite down to 80 nm by the Kramers-Kronig approach using EELS data from \cite{Tosatti1970}. 
\cite{Zubko1996} used the same method to obtain the optical constants of several solid hydrocarbons from the millimeter to the FUV down to 40 nm.  
\cite{Schnaiter1996, Schnaiter1998}  derived the optical constants through UV measurements of nano-sized carbon grains using the matrix-isolation technique in the 200 - 400 nm region. 
\cite{Furton1999} also obtained the optical constants of a hydrogenated carbon film by trial and error fitting in the range of 300 - 1000 nm. 
These values have a dispersion of between $\sim$10\% and 20\% at any given wavelength, and they compare closely to other published data for similar a-C:H materials, for example, \cite{Smith1984} between 253 nm - 900 nm, and for organic residues, for example, \cite{Jenniskens1993} from 115 nm to 1.7 $\mu$m. 
\cite{Jager1998} derived the optical constants  of carbonized cellulose samples using the Lorentz-oscillator fit method in the range of 200 - 500 nm based on reflectance measurements. This was done later for  gas-phase condensed soots from 120 nm to the mid-infrared \citep{Jager2008}. Few laboratory based determinations of optical constants of carbon analogs exist for $\lambda$ $<$ 120 nm.  \cite{Fink1984} measured the EELS spectra from 300 nm to 40 nm for amorphous carbons. \cite{Colangeli1993} measured the spectra of carbon samples down to $\sim$50 nm through transmission
or reflection on a gold-coated quartz window. 

This paper is organized as follows. In Sect. 2 we describe the experimental setups used in sample preparation and characterization of the carbonaceous materials. In Sect. 3 we present the UV and MIR spectroscopy of both materials. In Sect. 4 we describe the method used to derive the optical constants. In Sect. 5 we discuss the astrophysical implications of these measurements. Finally, in Sect. 6 we present our conclusions. 

\section{Experiments}
We studied two types of carbonaceous dust: a-C:Hs and soot samples. The a-C:Hs were produced using an R.F. plasma reactor at low pressures. Their typical structural unit is shown in \cite{Dartois2005}. Soot materials have been produced in an ethylene (C$_2$H$_4$) flame and provide samples dominated by a polyaromatic carbon skeleton (e.g., \cite{Carpentier2012}). While our a-C:H materials have a high H/C, our soot materials have low H/C.

Two types of substrates were used: MgF$_2$ windows and TEM grids (\textit{Ted Pella, Inc.}) covered with a nm-thick metallic or carbon membrane. Samples of MgF$_2$ were used to ensure UV-visible transmission measurements down to $\lambda$ = 120 nm. Samples deposited on the 3 mm TEM grids were used for transmission measurements from $\lambda$ = 120 nm to 50 nm.  The grids were 300 mesh Au or Cu substrates covered with a 2-3 nm layer of gold or platinum, or 10-25 nm of carbon. Manipulation of the TEM grids was done with high-precision tweezers, but more often than not, the fragile grid supports were bent and/or the thin membrane shattered, either during installation in the sample holder or during deposition of the film. We inspected the films with a binocular microscope at the Institut d\textsc{\char13}Astrophysique Spatiale and with microscopic images taken at the SMIS beam line at the SOLEIL synchrotron.

As a result of four experimental campaigns in 2014 and 2015 at the DISCO VUV beam line of the SOLEIL synchrotron, we found that while Au or Pt membranes (2 - 3 nm) had good transmittance in the VUV (T $\leq$ 30 \%), they were very fragile during manipulation and during film deposition. Most VUV measurements on these substrates showed that the membranes were damaged either partially or completely. Instead, the carbon membranes on TEM grids (T $\leq$ 20 \%) proved to be more robust. 
We discuss the effect of these substrates on the film transmission spectra in Appendix \ref{Ap3}. 

\subsection{Preparation of a-C:H films}

\begin{figure}[htbp]
\begin{center}
\includegraphics[width=90mm]{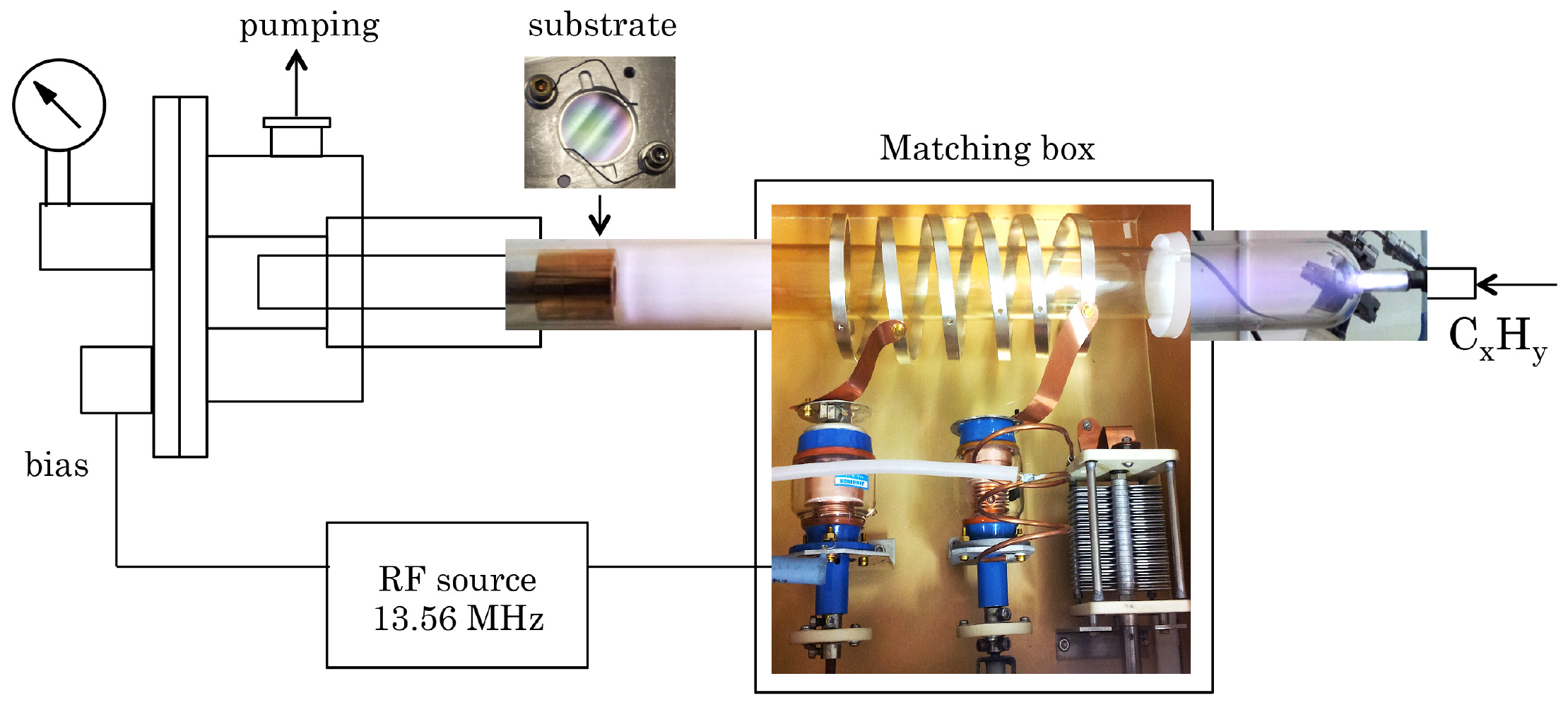}
  \captionof{figure}{\textit{SICAL- ICP} is a setup allowing plasma-induced chemical vapor deposition. It is comprised of a glass tube in which an inductively coupled plasma (ICP) on a continuum hydrocarbon gas flow is maintained by an R.F. source. The plasma envelopes the sample holder placed near the matching box.}
  \label{sical}
\end{center}
\end{figure}

\textit{SICAL-ICP} is an experimental setup developed for the purpose of producing hydrogenated amorphous carbon films, see Fig.~\ref{sical}. 
It consists of an inductively coupled radio frequency (R.F.) generator working at 13.56 MHz (\textit{Elite series}, MKS) with a dynamic range of 1 - 600 W. A background pressure of $<$ 0.01 mbar is maintained by a primary pump. The precursor hydrocarbon gas (CH$_4$) supply is manually controlled with a fine valve and deposits are made in continuous flow kept at $\sim$0.1 mbar, after passivation of the chamber with the precursor.

The R.F. discharge is operated in a glass tube (length = 1 m, $\phi_{ext}$ = 65 mm), allowing for the plasma to symmetrically envelop the sample holder. The R.F. power is inductively coupled to the plasma by means of an impedance matching network linked to a copper coil around the glass cylinder. The tuning box allows extracting a portion of the R.F. power coupled to the substrate holder acting as an externally applied bias. For most samples, the power was kept at 100 W, with 5 W reflected. 
The sample holder consists of a gold-coated copper block capable of holding 20 mm windows. The holder can be internally cooled by water, and the substrate can be installed by removing a mechanical cover. 
The estimated ion energies are on the order of 80 eV, thereby producing films with a hydrogen content up to 40\% \citep{Dworschak1990}. 
This R.F. source allows lower ion energy densities than with the 2.45 GHz source previously used by \cite{Dartois2005}, giving a-C:H films with higher hydrogen content due to the decreased preferential sputtering of the C-H bonds vs. the C-C bonds.  Lower pressures leading to higher ion energies also lead to sputtering of hydrogen. 
Samples are prepared at room temperature, although the substrate temperature has a strong influence on the film properties \citep{Dworschak1990}.  

To control the deposition rate for the films, we used the correlation between the optical depth of the 3.4 $\mu$m band and the thickness of the films (\textit{d}) for films with \textit{d} $>$ 100 nm, using their Fabry-Perot interference fringes that are due to internal reflections within the film itself.  We extrapolated this correlation for films with \textit{d} $<$ 100 nm (see Appendix \ref{Ap1} for details). As a first estimate of the refraction
index of these films in the optical, we used the relation in \citep{Godardthese} between the optical gap (E$_{04}$) and the refraction index measured for different a-C:H and a-C (amorphous carbon) films (e.g., \cite{Dischler1983, Lazar1998, Kassavetis2007}). For our a-C:H materials, we derived the Tauc band gap E$_{g}$ = 2.5 eV, or E$_{04}$ = 3.1 eV, and using the aforementioned relation, we estimate n$_0$ $\sim$1.4 - 1.8.  

\subsection{Preparation of soot films}
\begin{figure}[htbp]
\begin{center}
\includegraphics[width=75mm]{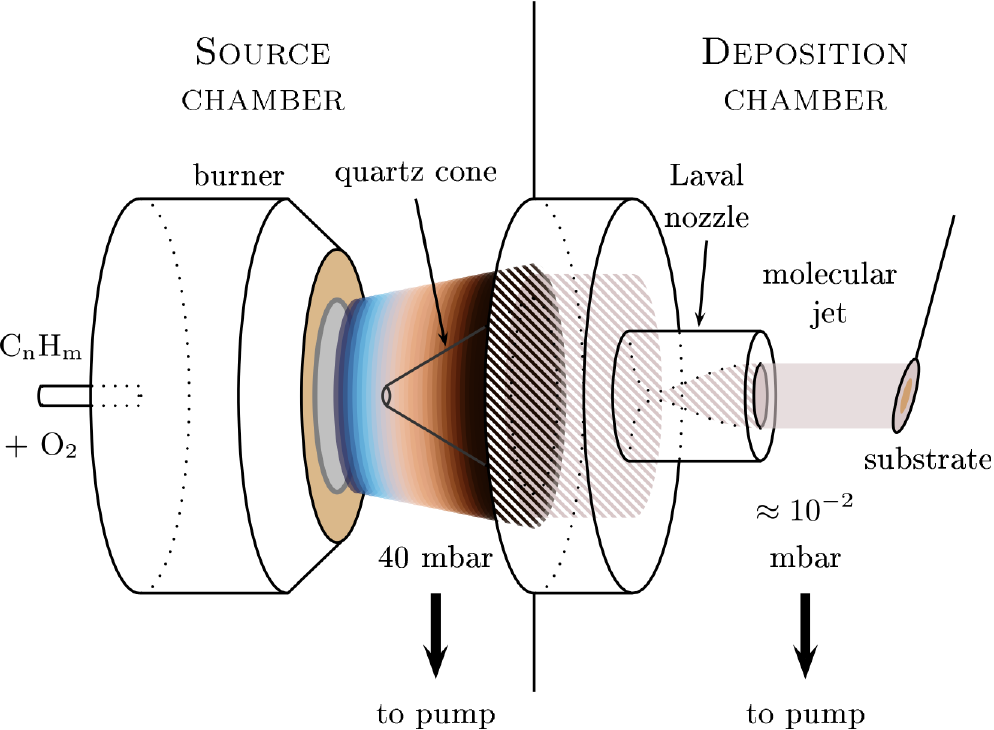}
\captionof{figure}{\textit{Nanograins}, an experimental setup for the preparation of soots. This scheme shows the burning region, the extraction cone
in quartz, and the gas flow region shaped by a nozzle where the soot samples are collected by impaction on a window \citep{Carpentier2012}.}
\label{fig:2} 
\end{center}
\end{figure}
Soot samples were produced at the Institut des Sciences Mol\'eculaires d'Orsay (France) using the setup \textit{Nanograins} \citep{Pino2008, Carpentier2012}. In the setup, whose scheme is shown in Fig. \ref{fig:2}, a flat burner (McKenna) provides flames of premixed hydrocarbon and oxygen gas. Such a laminar flame is a one-dimensional
chemical reactor offering a broad range of combustion conditions and sampling of byproducts. Here the pressure was maintained at 40 mbar by continuous pumping. The fuel used was ethylene (C$_2$H$_4$) premixed with oxygen before flowing through the burner at a controlled flow rate (4 liters/min), and a C/O ratio set to 1.05. An N$_2$ shield around the burning region was applied at a controlled flow rate of 3 liters/min. The byproducts were then extracted by a water-cooled quartz cone (hole diameter = 1 mm) inserted into the flame at 30 mm from the burner. The burned gas and soot flow through a nozzle into a chamber at 5$\times$10$^{-2}$ mbar. Soot particles were deposited on MgF$_2$/KBr/NaCl windows or Au/C/Pt membranes, inserted directly into the molecular flow. The thicknesses of the soot nanoparticle films varied in a range
of a few tens of nm to 1 $\mu$m, depending both on the production rate and the exposure time, to optimize the spectral response in the different wavelength domains from the IR to the VUV.

\subsection{Instruments for UV and IR spectroscopy}
Ultraviolet measurements of our carbon analogs were performed with three instruments. At the IAS we use a UV-Vis spectrometer ($\lambda$ = 210 nm - 710 nm) with a sub-nm resolution. It consists of a deuterium lamp, whose light is conducted by optical fibers and illuminates the sample. The transmitted signal is then collected and carried by optical fibers toward a grating spectrometer (Maya PRO UV, \textit{Ocean Optics}). For measurements in the vacuum ultraviolet, we use the APEX branch of the DISCO beamline at the SOLEIL synchrotron. The APEX chamber allows  measurements at $\lambda$ = 50 nm - 260 nm in windowless mode \citep{Giuliani2009}, with a beam size at the sample of $\leq$ 2 mm. From now on, we interchange between the terms VUV and FUV to refer to this  wavelength range. Finally, a spectrophotometer (Specord, \textit{Analytikjena}) allowed measurements on MgF$_2$ windows at $\lambda$ = 190 - 1000 nm. 

Infrared measurements for the thicker (\textit{d} $>$  300 nm) films deposited on MgF$_2$ windows were made using an evacuated FTIR spectrometer (Bruker Vertex 80V) equipped with a KBr beamsplitter and an (MCT) detector working in the 7000 to 400 cm$^{-1}$ (2.5 to 15 $\mu$m) spectral range at 1 cm$^{-1}$ resolution. 
To measure the infrared spectra of ultra-thin films (\textit{d} $<$ 100 nm) we used the infrared microscope available at the SMIS beam line at the SOLEIL synchrotron.  This microscope  benefits from high synchrotron brilliance in the 2.5 - 100 $\mu$m spectral range and a 10$\times$10 $\mu$m spot size on the sample. 

\section{Measurements} 
\subsection{UV spectra: a-C:H and soots}
\label{sec:UV}
\begin{figure}
\begin{center}
\includegraphics[width=90mm]{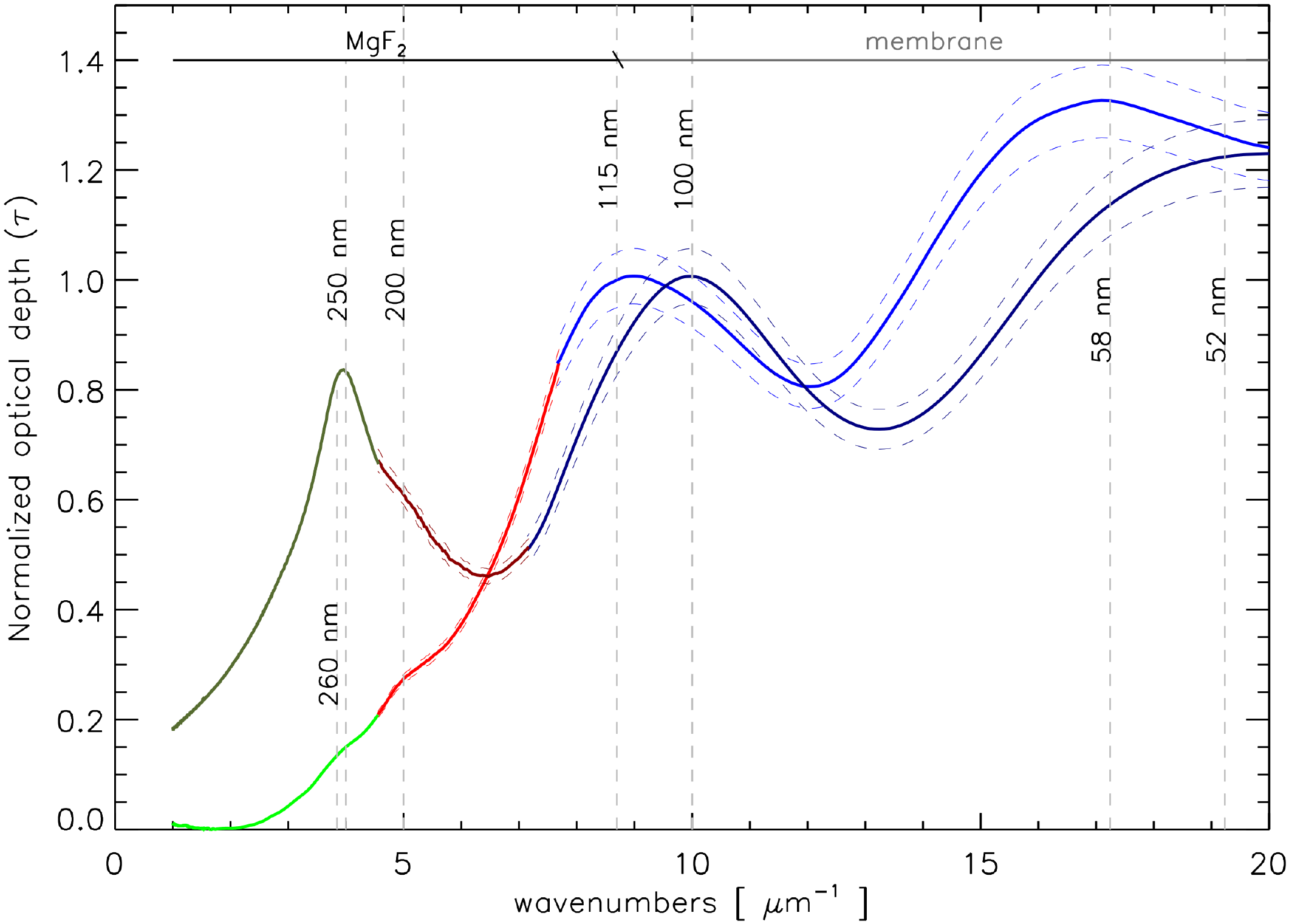}
  \captionof{figure}{VUV measurements of ultra-thin films:  a-C:H and soot optical depth spectra normalized to $\tau$($\sim$10 $\mu$m$^{-1}$).  a-C:H film,
  \textbf{{\color{LimeGreen} ---}}: MgF$_2$ (Specord), \textbf{{\color{red} ---}}: MgF$_2$ (DISCO), \textbf{{\color{blue} ---}}: TEM membrane (DISCO). Soot film, \textbf{{\color{OliveGreen} ---}}: MgF$_2$ (Specord), \textbf{{\color{Maroon} ---}}: MgF$_2$ (DISCO),  \textbf{{\color{RoyalBlue} ---}}: TEM membrane (DISCO). Dashed lines show measurement dispersion. }\label{fig4b}
\end{center}
\end{figure}
Transmission spectra were obtained by correcting each spectral measurement by an electronic offset and then dividing the spectra of the desired carbon film deposited on the substrate (MgF$_2$ or a membrane) by the spectra of the substrate alone. We assumed that there are no optical effects due to the substrates, which we further discuss in Appendix \ref{Ap3}.  The resulting normalized optical depth spectra for an a-C:H and a soot nanoparticle film are shown in Fig. \ref{fig4b}.

The UV-VUV spectra reveal the electronic structure of the carbon analogs, arising from transitions between threefold (sp$^2$ bonding) and fourfold (sp$^3$) bonding coordination, and any mixture of these two bondings (sp bonding is not considered here as it is a minor component in our samples). 
The optical gap reaches from states in the valence $\sigma$ and $\pi$ band to states in the conduction $\sigma^*$ and $\pi^*$ bands. 
The $\pi$ - $\pi^*$ bands due to sp$^2$ sites typically occur from 200 to 500 nm (2 - 5 $\mu$m$^{-1}$) and dominate the optical gap value.  The  $\sigma$ - $\sigma^*$ transitions peak below 200 nm (5 - 10 $\mu$m$^{-1}$). Both of these bands produce an absorption from the UV to the visible and infrared domain. 
The optical gap in amorphous or disordered carbon can extend to the IR induced by defect states. It can take a value of 0 eV for graphite, along a plane, and up to 5.5 eV for diamond. 

\begin{figure}[h]%
    \centering
    \subfloat[  ]{{\includegraphics[width=90mm]{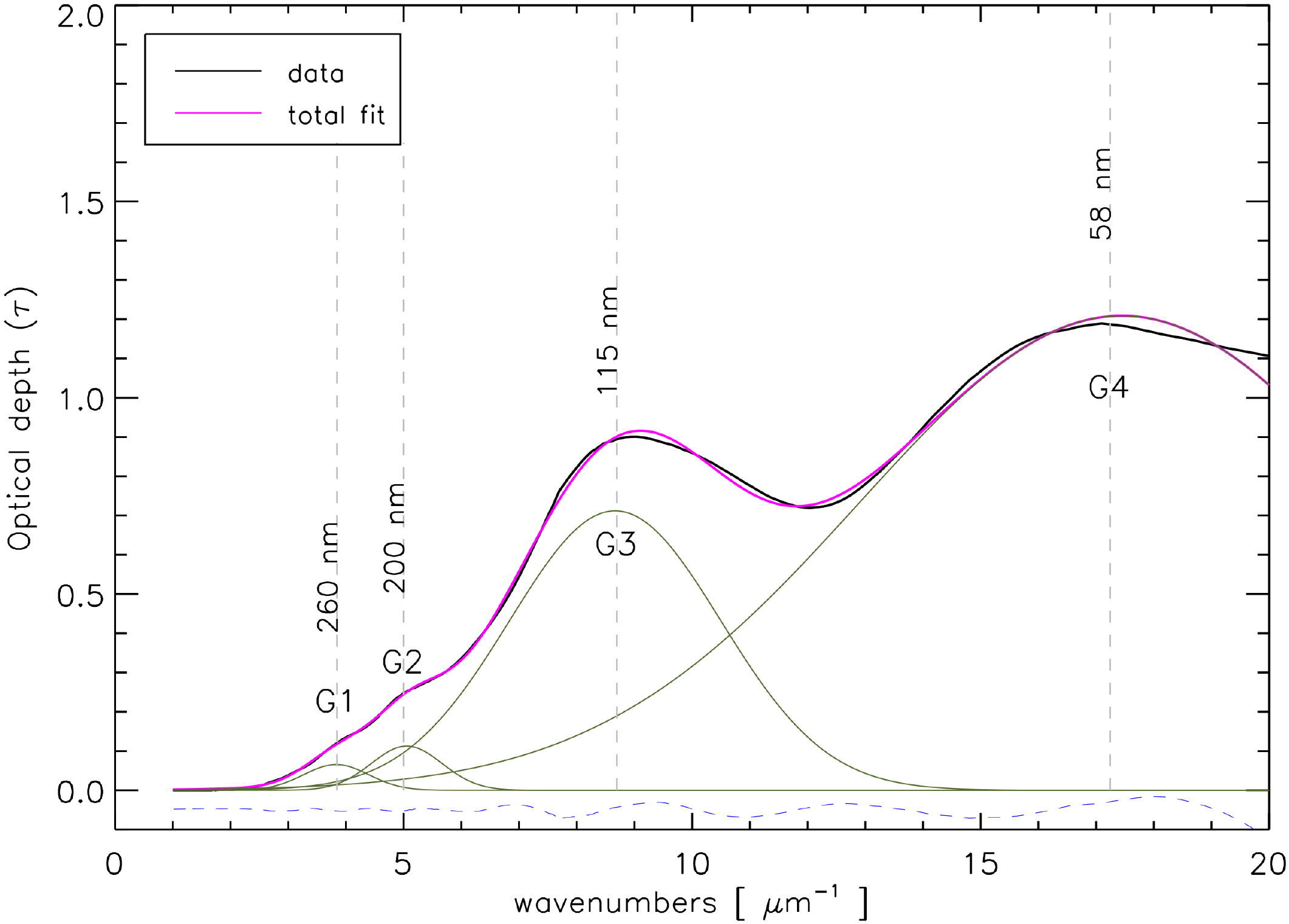} }}%
    \qquad
    \subfloat[  ]{{\includegraphics[width=90mm]{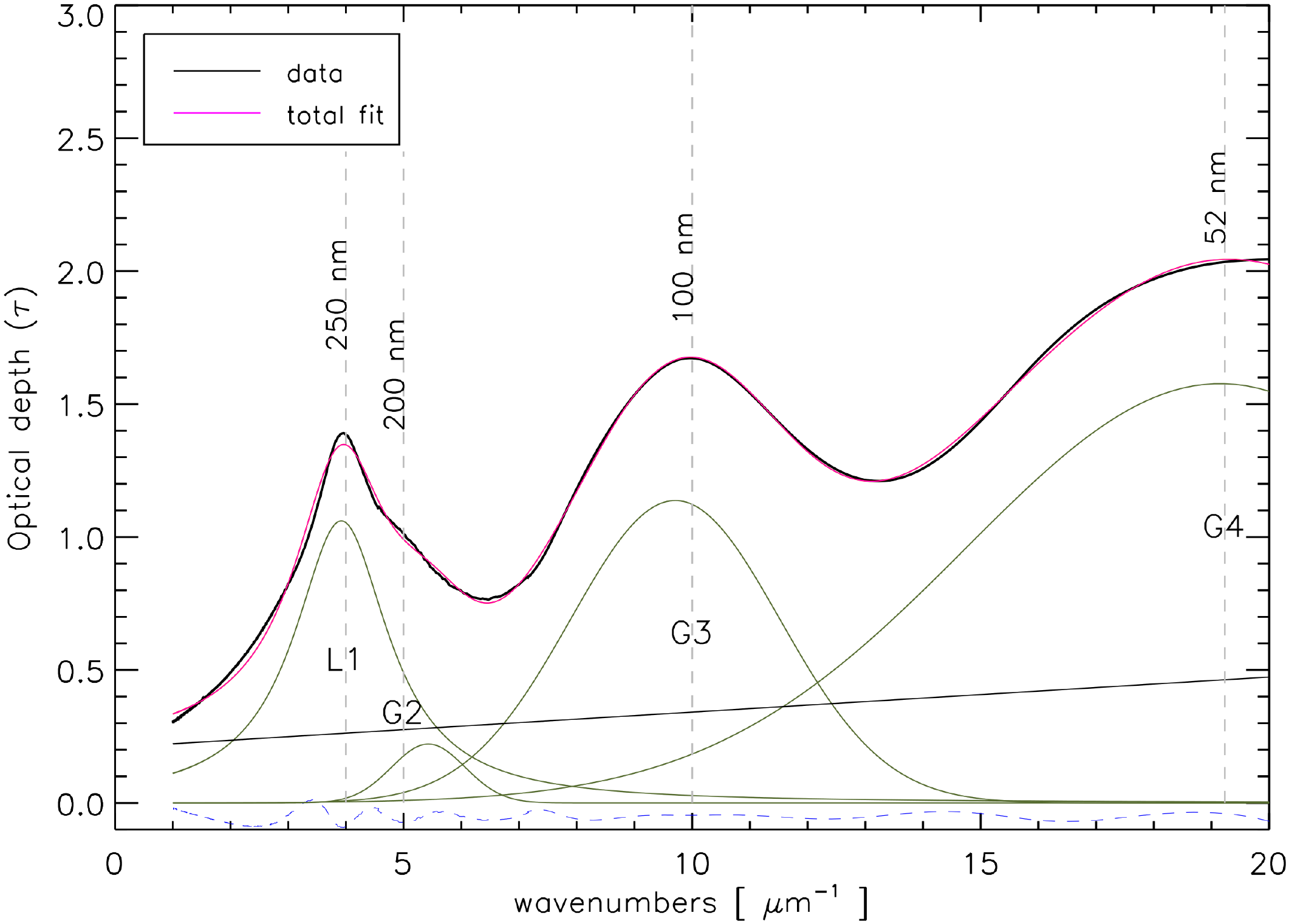} }}%
    \caption{VUV optical depth in the VUV of \textit{(a)} an a-C:H film and \textit{(b)} a soot nanoparticle film (described in the text). The deconvoluted spectral features are listed in Table \ref{tab1}.}%
    \label{fig:5ab}%
\end{figure}

An a-C:H sample of thickness $\geq$ 30 nm becomes optically thick in the VUV. However, these films are optically thin in the UV-Vis range, which allows us to discern weaker absorption features, such as the absorption bands at $\sim$4 - 5 $\mu$m$^{-1}$. The reconstructed VUV spectra of an ultra-thin a-C:H film (\textit{d} $\sim$20 nm ) and soot film (\textit{d} $\sim$50 nm) are shown in Figs. \ref{fig:5ab} (a) and (b), respectively. 
Fitting parameters are given in Table \ref{tab1}. 

For the a-C:H film, we identify four absorption features, fitted by Gaussian distributions. Two overlapping absorption peaks (G1, G2) occur at $\sim$4 and at 5 $\mu$m$^{-1}$. G1 is assigned to $\pi$ - $\pi^*$ transitions due to locally aromatic carbon clusters \citep{Robertson1987}. G2 is assigned to a superposition of carbon $\pi$ - $\pi^*$ transitions and/or \textit{n} - $\sigma^*$ transitions from oxygen contamination as discussed in \cite{Gadallah2011}. The third absorption (G3) peak at $\sim$9 $\mu$m$^{-1}$ (115 nm) in the far-UV is assigned to $\sigma$ - $\sigma^*$ transitions. A fourth peak (G4) is found at $\sim$16 $\mu$m$^{-1}$ (58 nm). A tentative identification of this fourth peak is given below. 
\begin{table}\centering
\centering
\captionof{table}{Best-fit parameters for VUV spectra} \label{tab1} 
\begin{tabular}{@{} cccc @{} }
\toprule
Function & Parameter [ $\mu$m$^{-1}$ ] & soot & a-C:H \\
\toprule
\multirow{3}{*}{Lorentzian (L1) } & area   & 3.33 &   \\ 
 & center  & 3.92 &  \\ 
& width & 2.01  & \\ 
\midrule
\multirow{3}{*}{Gaussian (G1)} & height & & 0.06   \\ 
& center   &   & 3.83  \\ 
& width &   & 5.64 \\ 
\midrule
\multirow{3}{*}{Gaussian (G2)} & height & 0.22 & 0.11   \\ 
& center   & 5.42  & 5.05  \\ 
& width & 6.37 & 6.13 \\ 
\midrule
\multirow{3}{*}{Gaussian (G3)} & height & 1.14  & 0.71 \\ 
& center & 9.71 & 8.67  \\ 
& width   & 1.81  & 1.83 \\ 
\midrule
\multirow{3}{*}{Gaussian (G4)} & height &  1.58 & 1.21   \\ 
& center   & 19.2  & 17.4  \\ 
& width & 4.41 & 4.55 \\ 
\bottomrule
\end{tabular}
\end{table}

The soot UV absorption spectrum was fitted by a Lorentzian, three Gaussians, and a linear function (slope $\sim$0.01 in energy units). The first peak (L1) is well-adjusted by a Lorentzian function peaking at 3.92 $\mu$m$^{-1}$ and assigned to $\pi$ - $\pi^*$ transitions. The relatively strong intensity of this feature points to a highly poly-aromatic material, close in shape but not in position to the interstellar UV bump at 4.6 $\mu$m$^{-1}$. A weak peak (G2) at 5.4 $\mu$m$^{-1}$ is assigned to  $\pi$ - $\pi^*$ transitions from carbon and/or \textit{n} - $\sigma^*$ transitions from oxygen contamination. A third broad peak (G3) occurs at 10 $\mu$m$^{-1}$ (100 nm) and a fourth peak (G4) is present at about 19 $\mu$m$^{-1}$ (52 nm). The linear function helps fit the soot opacity from 1 $\mu$m$^{-1}$  (while no linear function is required for the a-C:H fit).  

The position of peaks in the soot film are found at higher energies than for the a-C:H, with $\Delta$(L1-G1)= 0.19 eV, $\Delta$G2 = 0.45 eV, $\Delta$G3 = 1.62 eV, $\Delta$G4 =  2.47 eV. Two $\pi$-electronic transitions are found between 200 - 260 nm in both materials. For the hydrogen-richer a-C:H, the lower sp$^2$ carbon fraction decreases the $\pi$ density of states, which accounts for the weaker $\pi$ - $\pi^*$ band strengths. By taking the optical depth ratio of the L1 to G3 feature in the soot spectra (Fig. \ref{fig:5ab}a) and dividing it by the optical depth ratio of the G1 to G3 feature in the a-C:H spectra (Fig. \ref{fig:5ab}b), we find that the $\pi$ - $\pi^*$ resonance is $\sim$10 times stronger in soot than in a-C:H. 

Laboratory spectra of gas phase PAHs have shown their contribution to a FUV rise from a band with an extrapolated maximum at $\sim$17.7 eV (70 nm) \citep{Joblin1992}. Later on, \cite{Malloci2004}  computed the absolute photo-absorption cross sections of 20 PAHs in the VUV.  They found a good agreement both for the lower-lying bands of $\pi$ - $\pi^*$ character and for the collective single broad absorption peak at $\sim$17 - 18 eV, composed of $\sigma$ - $\sigma^*$, $\pi$ - $\sigma^*$, $\sigma$ - $\pi^*$ and Rydberg state transitions, with possible contributions from collective effects (e.g., plasmons). %
In graphite, in addition to the $\sigma$ and $\pi$ interband transitions, a band at 27 eV, measured in reflectance, was assigned to the collective excitations of the $\pi$ + $\sigma$ plasmon \citep{Djurisic1999}. %
In diamond-like carbon films (DLC), in addition to the $\sigma$ and $\pi$ interband transitions, a FUV contribution to absorption was detected using synchrotron ellipsometry, peaking at around 22 eV (56 nm, \cite{Franta2011}, Fig. 5). The transitions in this range are attributed by the authors to valence $\sigma$ to extended states transitions $\sigma$ - $\xi^*$. Figure 3 and Table 1 in \cite{Franta2011} depict the density of states for the corresponding band model. %
A marked double-peak structure was predicted by theoretical studies of the frequency-dependent dielectric permittivity in the optical to VUV domain of modeled carbon soot nanoparticles \citep{Moulin2008, Langlet2009}. The calculations predicted two peaks in the $\sim$20 - 30 eV range. The double-peak structure was suggested to arise from the polyaromatic units with a size distribution ranging from a few angstroms up to a few nm. In light of these findings, we tentatively assign the G4 band observed in both analogs to the features observed in this energy range in other carbonaceous materials. These transitions  involve excitations from $\pi$ and $\sigma$ states toward extended states above the $\sigma^*$ band. The nature of these excited states is still unknown and work is in progress to explore it. 

\subsection{Mid-infrared spectra: a-C:Hs and soots}

We measured the carbon analog thin films in the MIR with two objectives. The first is to correlate the integrated 3.4 $\mu$m band to the thickness of an a-C:H film to normalize the corresponding laboratory UV spectra to interstellar extinction curves. The second objective is to assess the homogeneity of the films and compare spectra from the same spatial region where the UV spectra is measured.  To this end, spectra with the SMIS FTIR microscope were taken every 0.2 mm across the ultra-thin a-C:H films deposited on MgF$_2$ and on TEM grids to check the reproducibility of depositions.   

\begin{center}
\includegraphics[width=90mm]{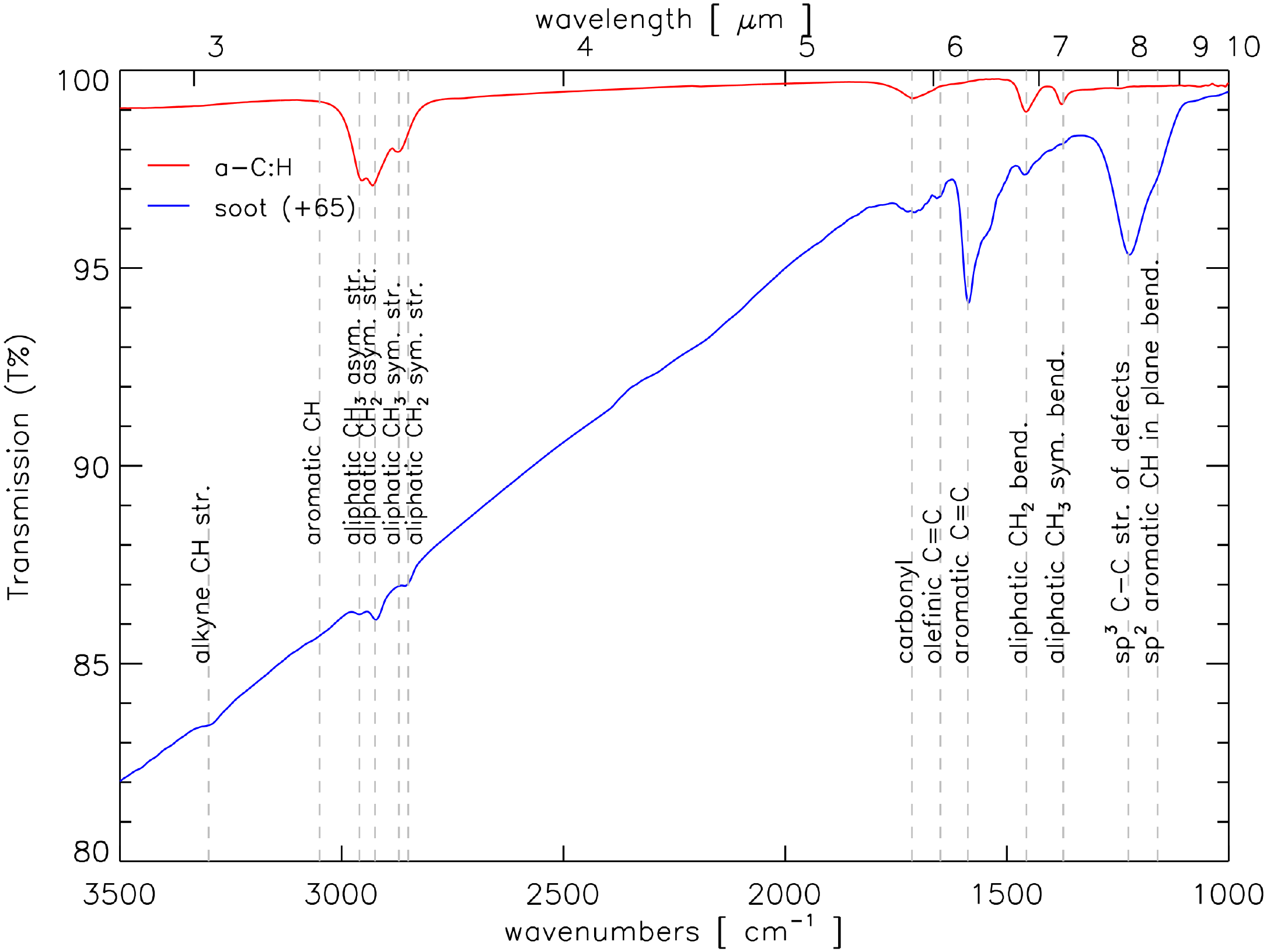}
  \captionof{figure}{Raw transmission spectra of an a-C:H film and a soot nanoparticle film (shifted by +65\% for comparison).}\label{fig:9ab}
\end{center}

Figure \ref{fig:9ab} shows the raw transmission mid-infrared spectra for an a-C:H and soot film. The continuum absorption is attributed to low-lying electronic transitions \citep{Carpentier2012}, while the superimposed bands are due to vibrational modes. 
The CH stretching bond region is seen at 2800 - 3100 cm$^{-1}$.  The bands between 1460 and 1380 cm$^{-1}$ correspond to the aliphatic CH$_n$ bending mode counterparts.  
The region between 1000 - 720 cm$^{-1}$ corresponds to out-of-plane aromatic CH bending modes that are non-detectable due to the low amount of sp$^2$ CH. 
Carbonyl is detected at $\sim$1718 cm$^{-1}$, mainly due to oxygen contamination. A weak OH band is present at $\sim$3400 - 3500 cm$^{-1}$.
\begin{figure}[h]
    \centering
    \subfloat[  ]{{\includegraphics[width=90mm]{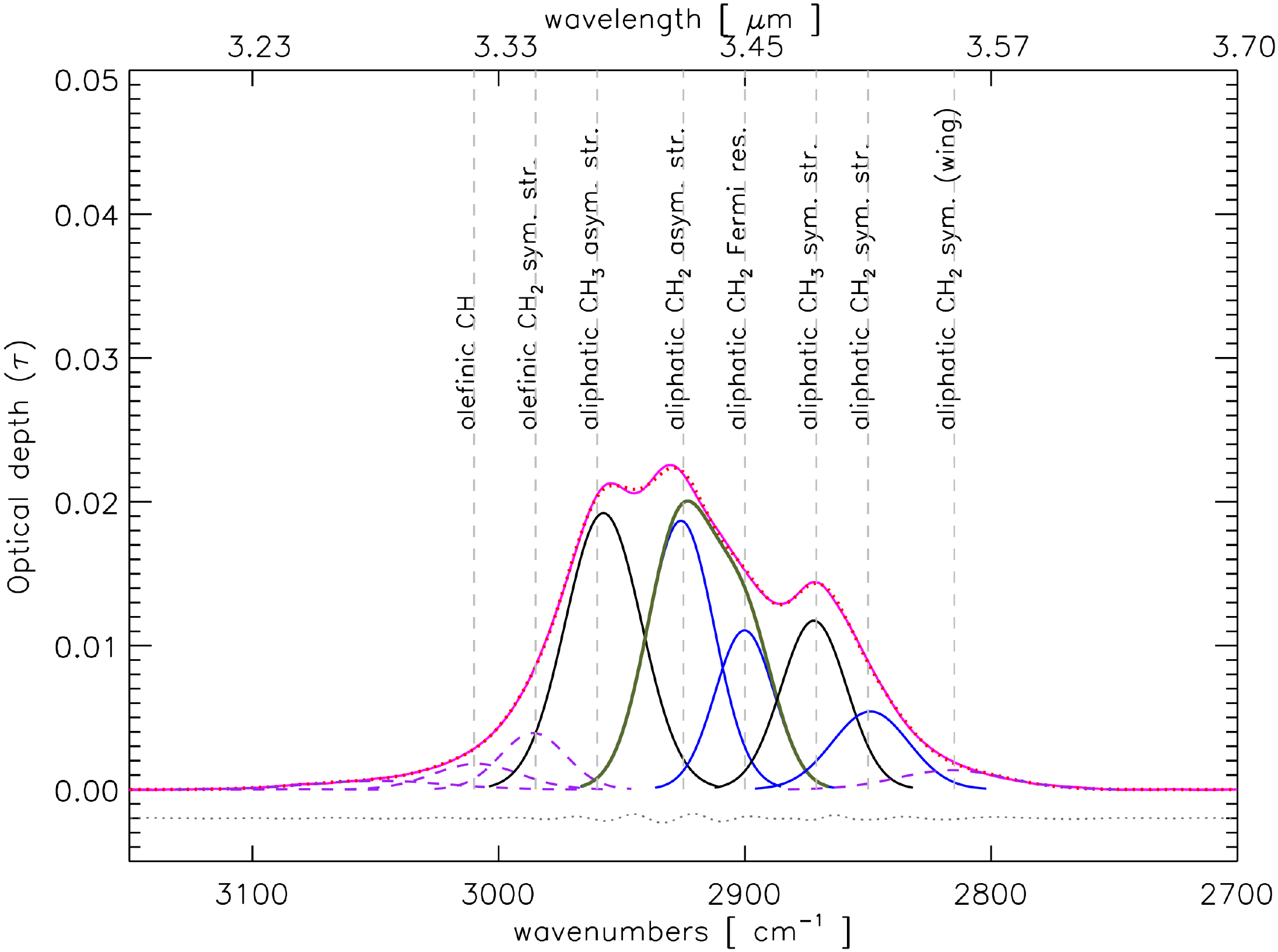} }}%
    \qquad
    \subfloat[  ]{{\includegraphics[width=90mm]{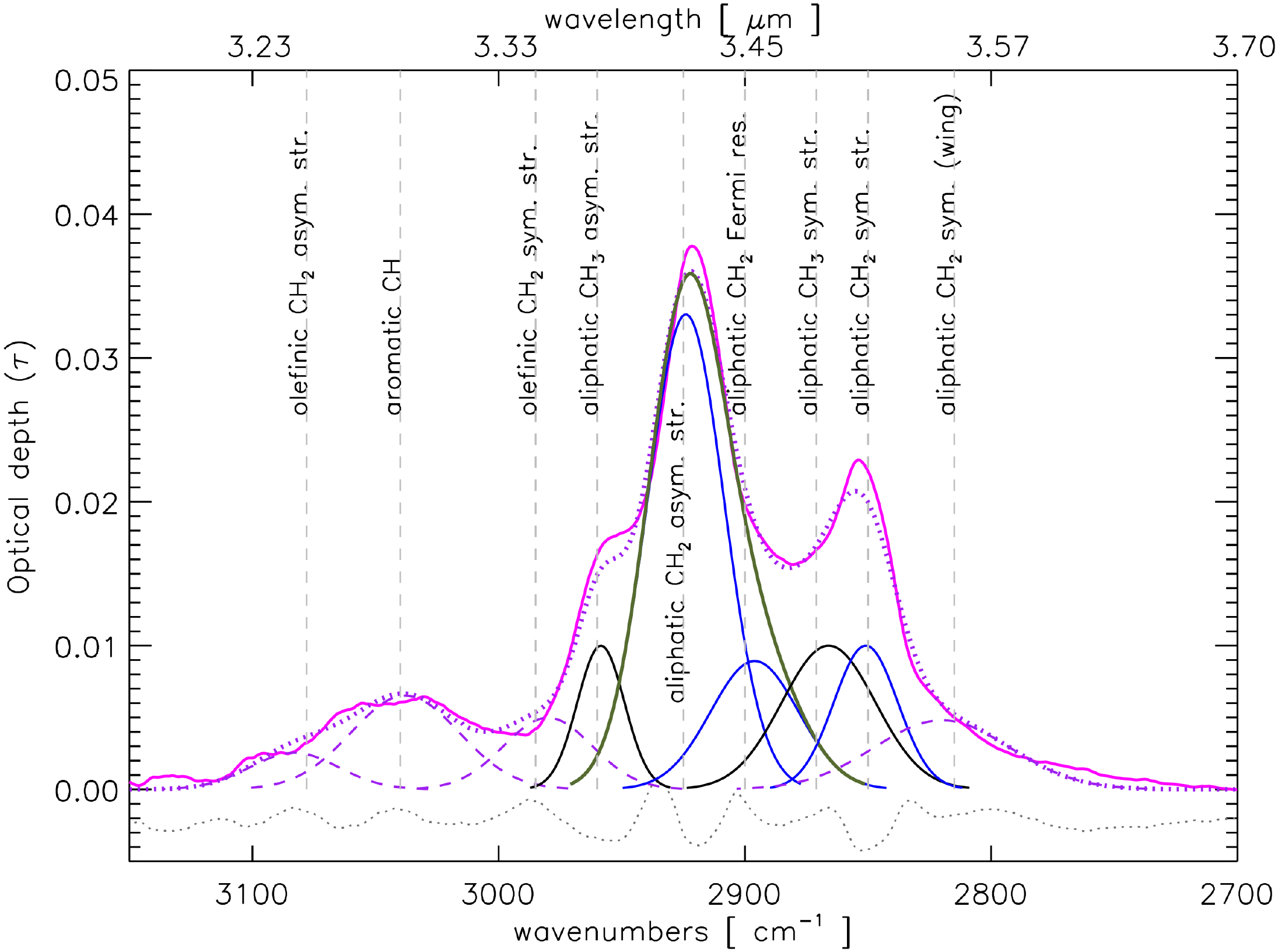} }}%
    \caption{Gaussian deconvolution of the 3.4 $\mu$m band of an \textit{(a)} a-C:H film and \textit{(b)} soot film. The black lines correspond to the -CH$_3$ groups, the blue lines to the -CH$_2$ groups. The fit residual is in gray.}%
    \label{fig:9ab1}%
\end{figure}

The intensity of the 3.4 $\mu$m band compared to the 6.2 and 8  $\mu$m bands is much lower for soots than for the a-C:H, where aliphatic features dominate.  For the a-C:H films we produce H/C $\sim$1, while for soots, the H/C ratio is lower ($\sim$0.01). 
The soot spectrum shows a 6.2 $\mu$m band (at 1584 cm$^{-1}$), due to the C=C sp$^2$ stretching modes. The 8.2 $\mu$m band (at 1225 cm$^{-1}$) is due to C-C (sp$^3$) stretching modes probably arising from defects, with a contribution at 1160 cm$^{-1}$ due to sp$^2$ aromatic CH bending modes. An alkyne CH stretching mode is present for soots at 3300 cm$^{-1}$ and is non-detectable for the a-C:H.  Band positions are coherent with previous laboratory soot infrared spectra \citep{Pino2008, Carpentier2012}. 

Figure \ref{fig:9ab1} (a) shows the deconvolution of the vibrational modes of the 3.4 $\mu$m band for a-C:H as in \cite{Dartois2004}.  The -CH$_2$ and -CH$_3$ groups dominate the spectrum of the band at 3.4 $\mu$m, but the shoulder at $\sim$3050 cm$^{-1}$ indicates the presence of olefinic and aromatic C-H bonds. 

The observed interstellar 3.4 $\mu$m band in the diffuse medium and the measured IR spectra of the produced a-C:H films agree
well \citep{Dartois2005, Dartois2007, Godard2010}. 
We used the oscillator strength of the observed modes to determine the aliphatic CH$_2$/CH$_3$ ratio of our a-C:H films \citep{Dartois2004b, Dartois2007}. The oscillator strength for the CH$_3$ antisymmetric mode, at 3.38 $\mu$m, A(a-CH$_3$) = 12.5 $\times$ 10$^{-18}$ cm/group, while for CH$_2$ antisymmetric + Fermi resonance at 3.377 $\mu$m, A(a-CH$_2$) = 8.4 $\times$ 10$^{-18}$ cm/group \citep{Dartois2004b, Dartois2007}. 
The calculated N(CH$_2$)/N(CH$_3$) ratio obtained from fitting the vibrational modes of the 3.4 $\mu$m band is 2.4 $\pm$ 0.3, in agreement with observations toward Cyg OB2 No. 12, which samples only diffuse ISM dust \citep{Pendleton2002},  where a ratio of 2.5 $\pm$ 0.4 is found, and to observations toward the infrared galaxy IRAS 08572+3915 \citep{Dartois2007}. 

Figure \ref{fig:9ab1} (b) shows the deconvolution of the vibrational modes of the 3.4 $\mu$m band for the soot nanoparticle film. From this, the calculated aliphatic N(CH$_2$)/N(CH$_3$) ratio = 7 $\pm$ 3. Using the oscillator strength for the aromatic CH vibrational mode at 3.289 $\mu$m, A(CH) = 1.9 $\times$ 10$^{-18}$ cm/group \citep{Joblin1994}, we find N(CH)$_{arom}$/N(CH$_{2,3}$)$_{ aliph}$ = 1.1 $\pm$ 0.3, in agreement with \cite{Carpentier2012}.  Using the band strengths from methyl-substituted PAHs \citep{Yang2013}, we find N(CH)$_{arom}$/N(CH$_{2,3}$)$_{ aliph}$  = 0.96 $\pm$ 0.2.

\section{VUV optical constants}
\label{sec:uvcon}
Optical constants were extracted from the transmission spectra of a-C:H films of varying thickness. For the soot film, a thickness $\sim$50 nm was estimated directly via AFM.  From the transmission spectra, we obtained first-order values for the imaginary index of refraction, k($\bar\nu$), according to the Beer-Lambert law,
\begin{equation}
{\rm
k  \sim \frac{\tau (\bar\nu) }{ 4 \pi \bar\nu d}, 
}
\end{equation}
where ${\rm \tau}$ is the optical depth, ${\rm \nu}$ is in wave
number units, and $d$ is the sample thickness. Before this, a baseline for the interference fringes due to multiple internal reflections was removed. By fitting the absorption in \textit{k} by a series of Gaussian profiles, \textit{k} was then extrapolated to 0 eV and to higher energies, to apply the Kramers Kronig relations, closure relations linking the real and imaginary parts of the complex dielectric index of refraction  \citep{Bohren1998}. 
To derive the real part of the complex index of refraction from our absorption data, we used the subtractive Kramers-Kronig (SKK) method \citep{Warren1984}. In this method, we fixed n($\nu_0$), known as the anchor point, a known value for the index of refraction at a reference frequency far from any strong absorption in the material considered, generally in the NIR-visible. 
 SKK is the preferred method if \textit{k} is only known over a limited interval and the integrals are evaluated numerically. 
For the real index, 
 \begin{equation}
 {\rm
n(\nu) = 1 + \frac{2}{\pi}P\int_{0}^{\infty}\frac{\nu'k(\nu')}{\nu'^2-\nu^2}d\nu' 
}
,\end{equation}
 where P denotes the Cauchy principal value of the integral. Inversely, for the imaginary index,

\begin{equation}
{\rm
k(\nu) =  \frac{2 \nu}{\pi}P\int_{0}^{\infty}\frac{n(\nu')}{\nu^2-\nu'^2}d\nu' 
}
.\end{equation}

For improved accuracy we used the formulation in \cite{Trotta1996}, which reduces the error by using a finite wavelength integration range,

\begin{equation}
{\rm
n(\nu) = n(\nu_0) + 
\frac{2}{\pi}P\int_{\nu_1}^{\nu_2}\bigg[\frac{k(\nu')\nu'-k(\nu)\nu}{\nu'^2-\nu^2}-\frac{k(\nu')\nu'-k(\nu_0)\nu_0}{\nu'^2-\nu_0^2}\bigg]d\nu' 
}
.\end{equation}

\begin{figure}[ht]%
    \centering
    \subfloat[  ]{{\includegraphics[width=80mm]{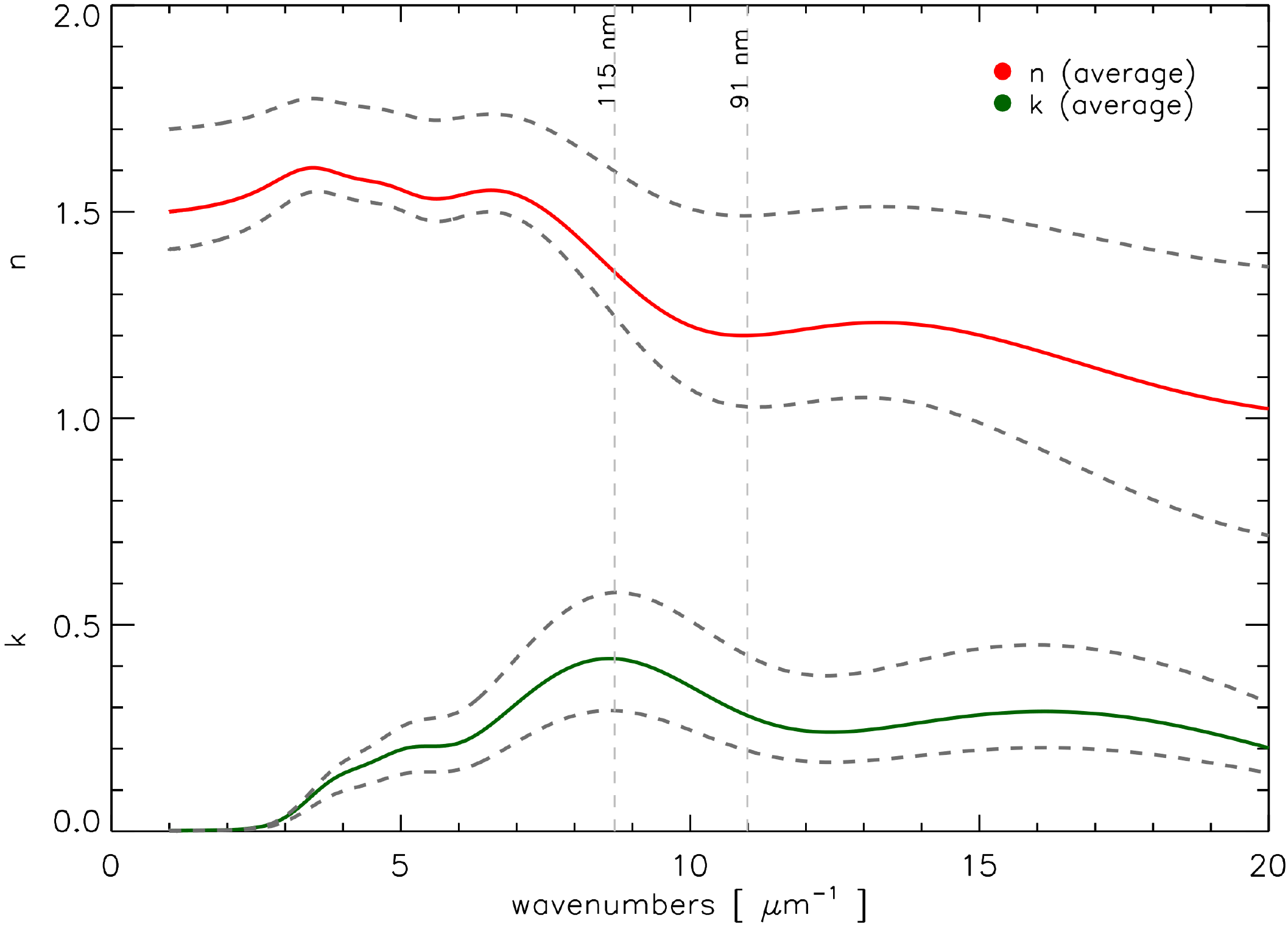}}}%
    \qquad%
    \subfloat[  ]{{\includegraphics[width=80mm]{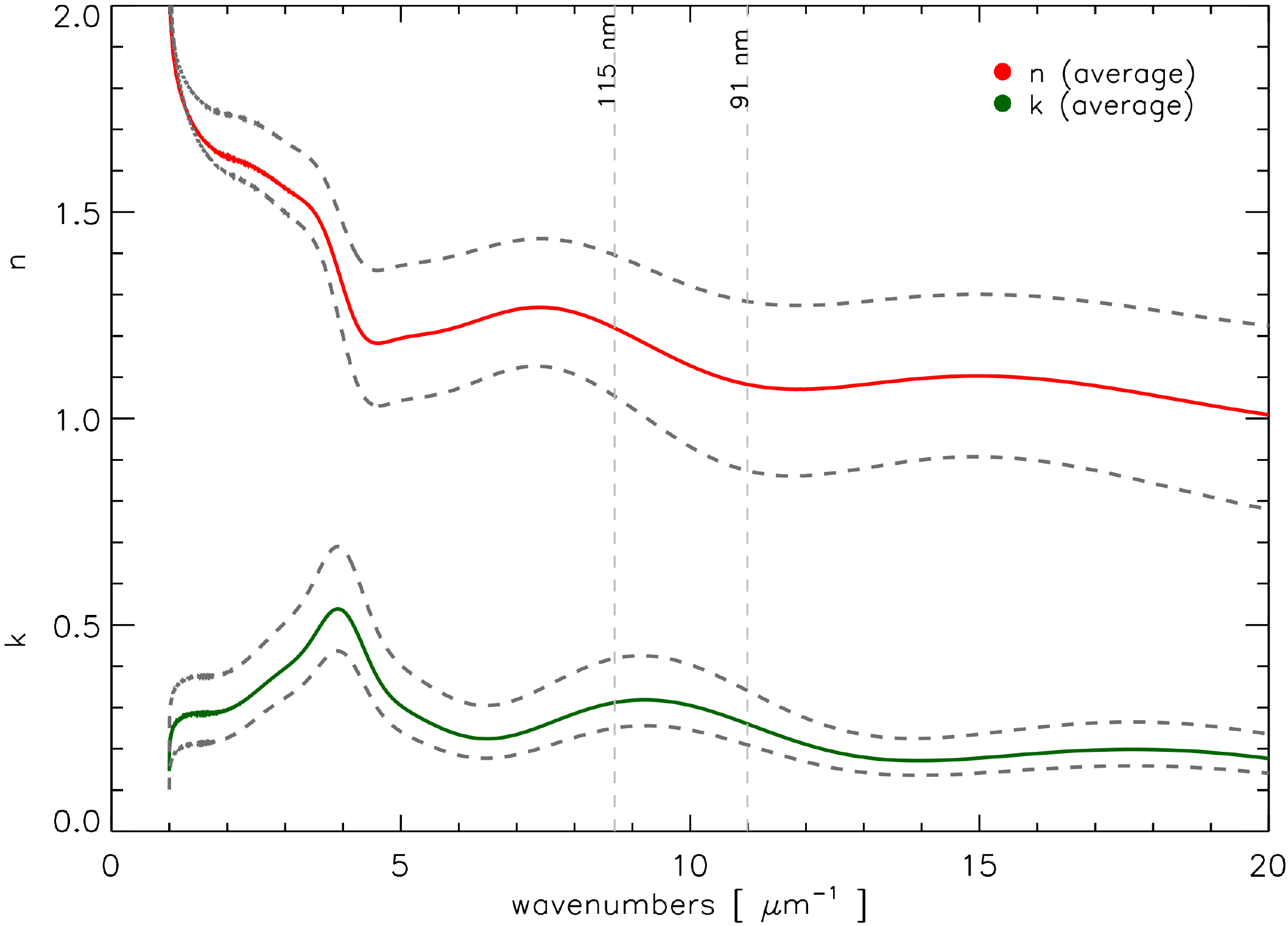}}}%
    \caption{Derived optical constants for \textit{(a)} an a-C:H film and \textit{(b)} a soot nanoparticle film, measured down to $\lambda$ = 50 nm. The dotted lines represent the compounded error bars dominated by uncertainty in the thickness and anchor points (n$_0$).}%
    \label{fig:nk}%
\end{figure}

For a-C:H we took n$_0$ = 1.5 as an anchor point at 1 $\mu$m. For soot we took n$_0$ = 1.7 \citep{Schnaiter1998}. The Kramers-Kronig numerical integration calculates the desired $n$ value over the input energy range. Subsequently, the complex index of refraction values were used to numerically simulate a synthetic transmission value at each wavelength. For this, we used the model of \cite{Swanepoel1983}, who derived the expression for the transmission of a thin absorbing film on a transparent substrate (MgF$_2$ in our case). We constructed a real index template for a substrate made of MgF$_2$ from 120 - 1000 nm. 
For $\lambda$ = 50 - 120 nm an inversion was made, keeping a constant \textit{n} value, but eventually, a more sophisticated scattering model should take into account the effects of a non-transparent substrate.  The difference between the synthetic and measured transmission was calculated at each point until the difference between the observed and synthetic spectra was minimized, which occurred after a few iterations. 

The optical constants derived for the laboratory a-C:H and soot films are shown in Fig. \ref{fig:nk} (a) and (b). 
For a-C:H the systematic uncertainty is mostly due to the thickness estimated from the n*thickness vs. 3.4 $\mu$m correlation (see App. \ref{Ap1}), where uncertainties in the thickness of the ultra-thin film go up to $\sim$50\% for a-C:H. 
The uncertainty for the soot film thickness is as large as $\sim$25\% (estimated from AFM measurements).    
The imaginary index of refraction reflects the major absorptive features seen in the laboratory spectra, that is, two FUV bumps  for both materials at $\ge$ 10 $\mu$m$^{-1}$ and a Lorentzian peak at $\sim$4 $\mu$m$^{-1}$ for the soot. The uncertainty in the thickness measurement dominates the systematic uncertainty, demonstrating the experimental difficulty of measuring ultra-thin films in the VUV. As in \cite{Franta2011}, who measured the optical properties of DLC films by ellipsometry in the 5 - 30 eV range, the present study shows that direct optical measurements (e.g., in transmission) are necessary to remove ambiguities in calculating optical constants in the VUV, especially concerning the contribution of the band at $\sim$22 eV, which are expected by the models described in Sect. \ref{sec:UV}. 

\section{Astrophysical implications}
\begin{figure}[h]%
\begin{center}
\includegraphics[width=85mm]{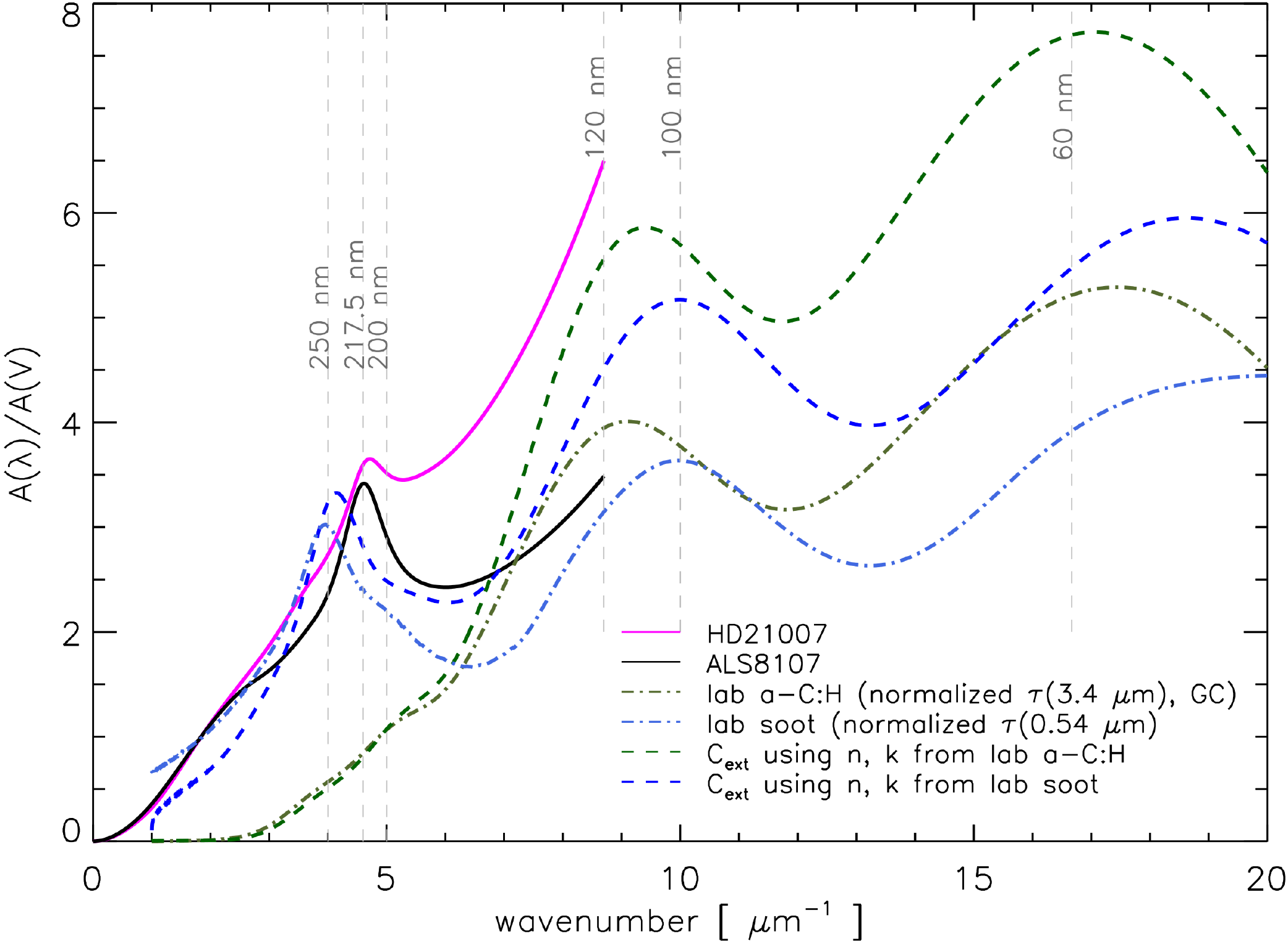}
\captionof{figure}{Normalized extinction, A($\lambda$)/A(V), of representative lines of sight from \cite{Fitzpatrick2007} (solid purple and black). Laboratory spectra for an a-C:H re-normalized to the optical depth of the observed 3.4 $\mu$m band (dotted green) and renormalized laboratory soot spectra to its V band optical depth (dotted blue). Calculated extinction curves from laboratory derived optical constants for an a-C:H (dashed green curve) and a soot (dashed blue). }
\label{fig:fmall}     
\end{center}
\end{figure}
\cite{Fitzpatrick2007} analyzed 328 Galactic interstellar extinction curves from the UV to the IR. Figure \ref{fig:fmall} shows two selected extinction curves (HD21007 and ALS8107) which may represent a-C:H and soot components. The laboratory a-C:H  spectrum is scaled to its laboratory 3.4 $\mu$m optical depth and to the expected observational ratio of 250 toward the Galactic center (Fig. \ref{fig:fmall}).  Observations of the optical depth of the 3.4 $\mu$m band have been compared to the variations in visual extinction,  A$_V$, toward Galactic sources and diffuse ISM lines of sight, and a correlation between these two has been found \citep{Sandford1995, Godardthese}. A ratio of A$_V$/$\tau_{3.4}$ $\sim$250 has been estimated for lines of sight of the local ISM, while observations toward the Galactic center give a ratio of A$_V$/$\tau_{3.4}$ $\sim$150 \citep{Sandford1995, Rawlings2003}, although the difference is probably lower \citep{Godard2011}.  The soot optical depth spectrum (Fig. \ref{fig:fmall}) is scaled to its optical depth in the V band (551 nm). 
 
We used the laboratory optical constants obtained in Sect. \ref{sec:uvcon} to calculate the extinction, Q$_{ext}$, that is due to a distribution of grain sizes, computed using the Mie formulae \citep{Bohren1998}. Our primary objective is to illustrate the effect of the laboratory optical constants from this study on extinction curves. As a first attempt, we took into account a simple size distribution for \textit{Very Small Grains} (VSGs) from \cite{Desert1990}, that is, 
\begin{gather}
{\rm
n(a) \propto a^{-p}
}
\end{gather}
where p = 2.6, and the grain sizes varied between a$_{min}$ = 1 to a$_{max}$ = 15 nm. 
The total extinction is then calculated as\begin{equation}
{\rm
C_{ext}  [cm^{-1}] \propto \int_{a_{min}}^{a_{max}} \pi a^2 n(a) Q_{ext} \,da 
}
.\end{equation}
This analysis is limited to spherical, homogeneous, optically isotropic particles, and the resulting extinction is given by the green and blue dashed curves in Fig. \ref{fig:fmall}. These optical constants should be implemented in more sophisticated extinction models. 
The calculated extinction shows that both materials present a steep FUV rise. However, the position of the FUV rise is primarily dictated by the position of the maximum of the $\sigma$ - $\sigma^*$ band. For a-C:H band maximum positions occur at lower energies than for soots, which dominate the steep FUV rise. 
The UV peak of the calculated extinction curve calculated for laboratory soots is closer in position to the average galactic extinction UV bump at 4.6 $\mu$m$^{-1}$ \citep{Fitzpatrick2007}.   

\cite{Draine1984} suggested that crystalline graphite particles could be a carrier of the UV bump, but due to the difficulty of condensing graphite in space and the need for specific size distributions for spectral agreement  \citep{Czyzak1981}, other carriers for this bump were sought. \cite{Mennella1996, Mennella1999} and later \cite{Gadallah2011} showed that an increase of the  sp$^2$ carbon fraction as a result of UV irradiation of their carbon materials could give rise to the UV bump. Considering the mid-infrared and UV features, their samples are more comparable to our soot than to our a-C:H sample. 
\cite{Li2008} used buckyonions to reproduce observed interstellar extinction curves. \cite{Jager1998} synthesized carbon analogs by pyrolizing cellulose materials and later by laser ablation of graphite \citep{Jager2008}. However, their \textit{\textup{onion-like}} carbon nanoparticles did not show a strong UV bump. 
Polycyclic aromatic hydrocarbons (PAH) were proposed to simultaneously contribute to the 217.5 nm bump and the FUV rise \citep{Malloci2004,Joblin2009}.  \cite{Steglich2010, Steglich2013} suggested that PAHs with sizes around 50 - 60 carbon atoms per molecule could be responsible for the UV bump through a heterogenous distribution of sizes and not through single isolated species. \cite{Cecchi2008} showed that these two spectral features and their variations can be accounted for by mixtures of PAHs in different ionization states.  More recently, hydrogenated amorphous carbon grains, believed to be more easily formed in carbon star atmospheres, have been considered as potential carriers of the UV bump, particularly when irradiated by UV photons \citep{Mennella1996, Mennella1997, Gadallah2011}. 

\cite{Li1997} modeled interstellar extinction by using a three-component dust model: silicates, amorphous carbons, and PAHs. Recent extinction models including only two components (silicates and a-C/a-C:H) \citep{Jones2012a, Jones2012b} have found good agreement with previous work. More importantly, they account for the FUV rise and the UV bump only by means of the size-dependent properties of the carbonaceous component. The calculated extinction curves (using a a size distribution of carbon dust grains characterized by our optical constants) show that soot-like nanograins could be an important contributor to the UV bump, while a-C:Hs can contribute more significantly to the steep FUV rise. However, other dust components (like silicates) should be considered in a complete model of the observed extinction.

\section{Conclusions}
We have described the preparation of laboratory a-C:H and soot nanoparticle analogs to dust found in the ISM. We characterized thin films (\textit{d} $<$ 100 nm) of these materials to measure their transmission spectra in the $\lambda$ = 1 $\mu$m to 50 nm range. These curves can be decomposed into absorption features involving transitions from states in the valence $\pi$ band to states in the conduction $\sigma$ band around 5 and 10 eV, the position and relative intensity varying with the sp$^2$ content of the material. The description of the measurements required
adding another band to explain the absorption, which peaked in the 16 - 19 eV range. This band is predicted by calculations and is present at varying levels in other carbonaceous materials.
It involves interband transitions to extended states.

By combining UV measurements to IR measurements, we scaled the UV contribution to the 3.4 $\mu$m band to compare the observed UV extinction to our laboratory a-C:H measurements. We derived the optical constants for a laboratory a-C:H and soot from several sets of transmission measurements at different thicknesses. We used these optical constants to calculate extinction curves that show that a-C:Hs contribute to the steep FUV rise, while soots are suitable carriers of the UV bump. Astronomical spectra of dust shortward of the Lyman limit are currently lacking, mostly due to the absorption lines of hydrogen and helium, which have large FUV and extreme-UV (EUV) effective absorption cross sections \citep{Hawkins1991}.  However, in dusty regions (where the opacity due to the gas component may be lower than the dust opacity), the laboratory FUV spectral features of carbon dust may be observed. 
\\ 

\vspace{0.05cm}
{ \footnotesize
\textit{Acknowledgments.} We thank Christophe Sandt (SOLEIL) for his help with the infrared microscope at the SMIS beam line. VUV measurements were made at the DISCO beam line of the SOLEIL synchrotron radiation facility (projects: 20141074 \& 20130778). This work has been supported by the French program ``Physique et Chimie du Milieu Interstellaire'', the ANR COSMISME project (grant ANR-2010-BLAN-0502). L.G. acknowledges support from the CNES post-doctoral fellowship program.  }
\bibliography{vuv}

\appendix
\section{Thickness estimation of an a-C:H film}
\label{Ap1}
The optical depth, $\tau$, obtained from transmission spectra, is equal to 
\begin{equation}
{\rm
\tau = \alpha d
}
,\end{equation}
where $\alpha$ is the absorption coefficient, and $d$ is the film thickness.
The absorption coefficient $\alpha$ is related to the imaginary part of the complex index of refraction, $k(\nu)$ by
\begin{equation}
{\rm
\alpha(\nu)= \frac{4 \pi \kappa(\nu)}{\nu}
}
,\end{equation}
which gives a first iteration of the $\kappa$($\nu$) index. From the Snell-Decartes law, we know that the optical path difference $\delta$  between a beam traversing the film and being reflected over each face of the film is 
\begin{equation}
{\rm
\delta = 2 n_0 d cos(\theta_f)
}
.\end{equation}
The interference fringes of the film will have a maximum when
\begin{equation}
{\rm
\delta = m \lambda  = \frac{m+1}{\Delta\sigma}
}
,\end{equation}
where $\Delta\sigma$ is the interfringe in wave numbers. Combining the previous two equations, we find that the thickness of the film is given by
\begin{equation} \label{eq12} 
{\rm
d = \frac{1}{2 n_0 \Delta\sigma}
}
,\end{equation}
where n$_0$ is assumed to be wavelength invariant, which is reasonable for visible and near-infrared wavelengths. 
\begin{figure}
\begin{center}
\includegraphics[width=85mm]{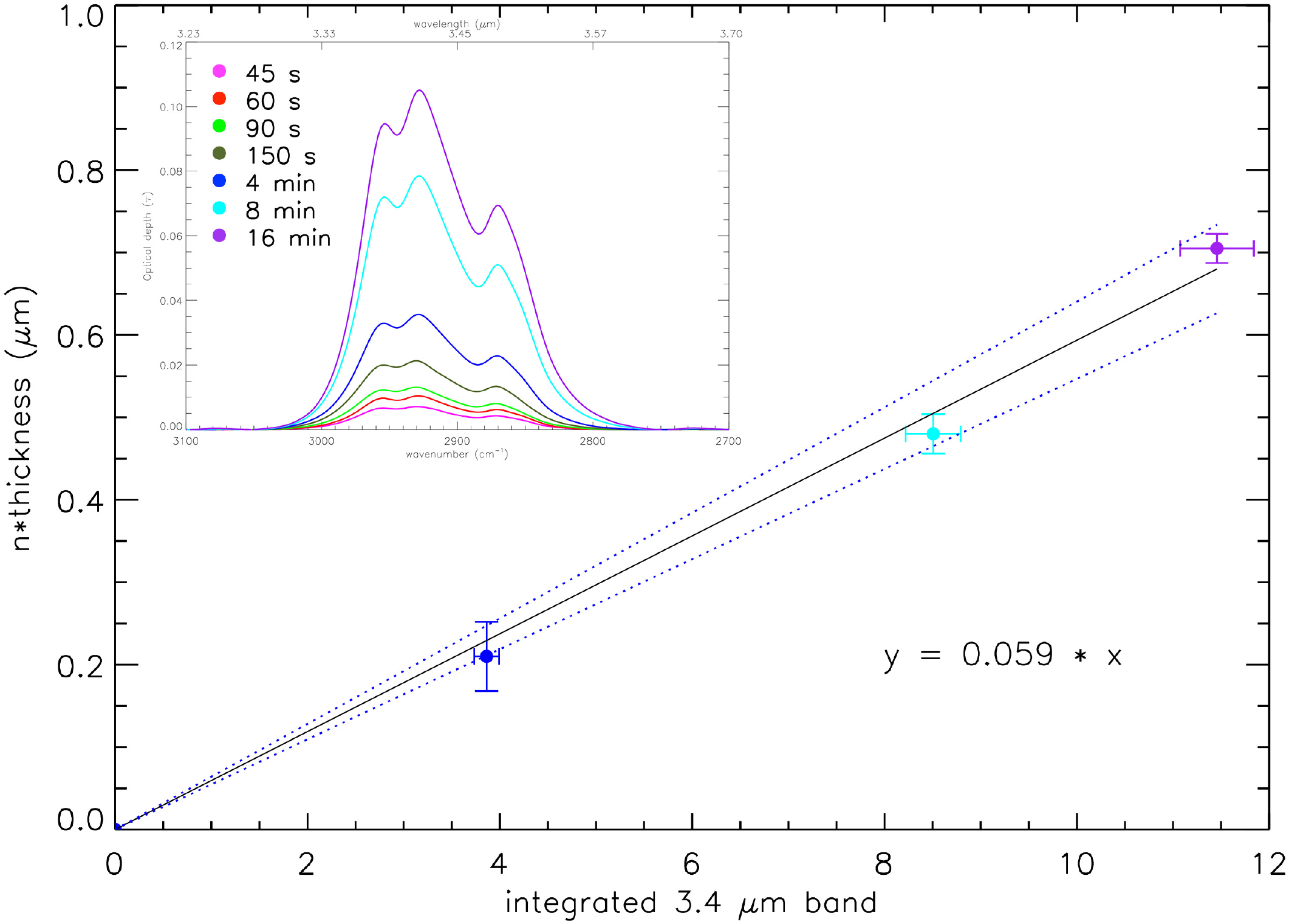}
\captionof{figure}{Correlation between n*thickness of films (\textit{d} $>$ 300 nm) vs. the integrated 3.4 $\mu$m band area.  This is used to estimate n*thickness for the ultra-thin films (\textit{d} $<$ 100 nm) using their integrated 3.4 $\mu$m values. The inset shows the 3.4 $\mu$m band of a-C:H films produced with different deposition times.}
\label{fig:nep}   
\end{center}
\end{figure}
Measurements in the mid-infrared (from 400 to 8000 cm$^{-1}$) give a limit for the detectable half interfringe of $\sim$200 nm. This means that the infrared interfringe of ultra-thin films is not measurable. We used the correlation between the integrated 3.4 $\mu$m band and n*thickness (from the interfringe measurements of the thicker films, from Eq. \ref{eq12}) to estimate the thickness for the ultra-thin films. Figure \ref{fig:nep} shows this correlation fitted by a linear function y = m*x, where m = 0.059 $\pm$ 0.01, along with 3$\sigma$ error bars. 
\section{Substrates in the VUV}
\label{Ap3}
%
\begin{center}
\includegraphics[width=85mm]{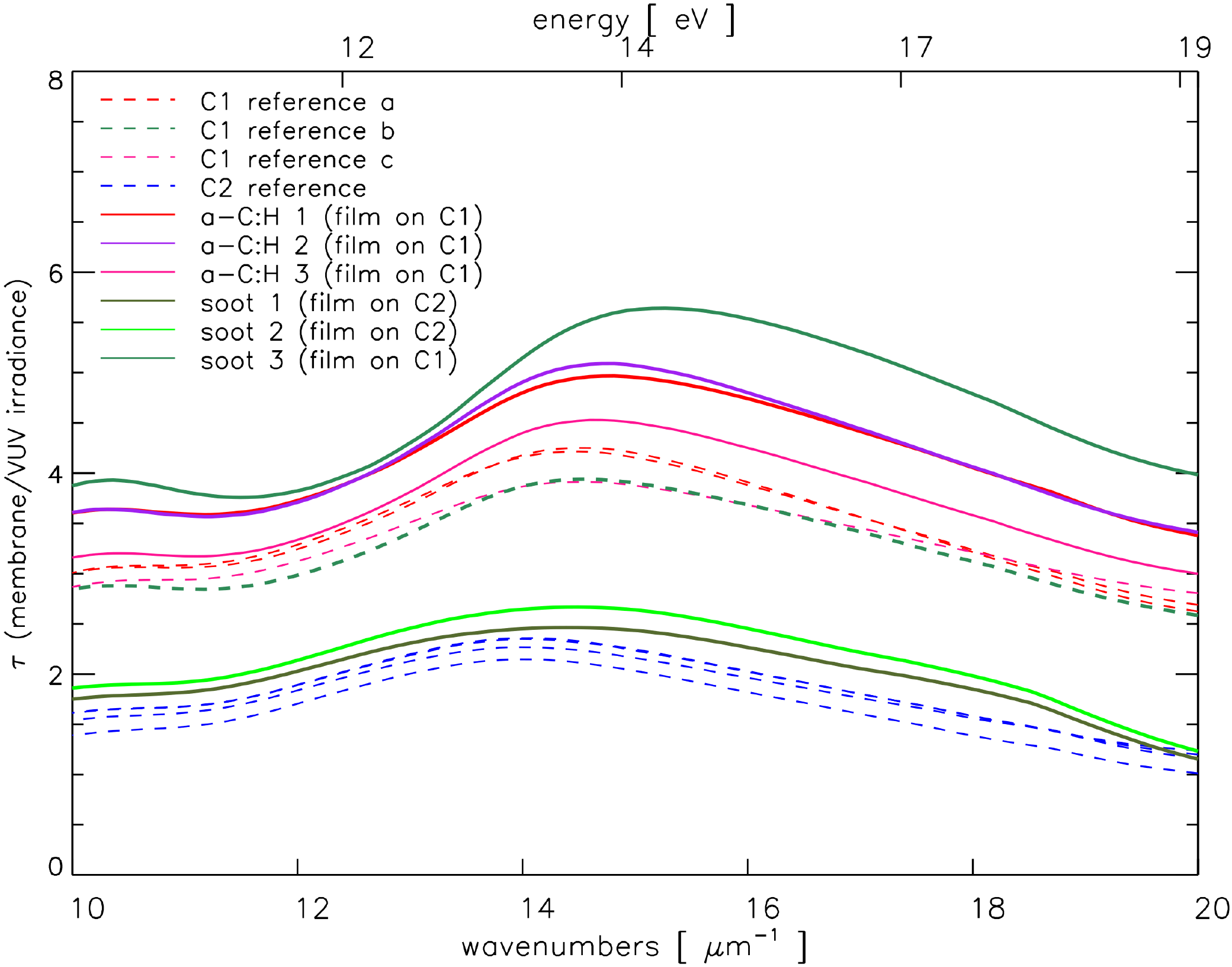}
\captionof{figure}{Transmission spectra of different carbon membranes (10 - 25 nm thick) used for VUV measurements (dashed lines), and carbon membranes + a-C:H or soot deposits (solid lines). }\label{fig:cmemb}     
\end{center}
To study the influence of the substrate on the transmission spectra of a-C:H and soot films in the VUV, we studied the spectral dispersion of the substrates and the films deposited on such substrates, seen in Fig. \ref{fig:cmemb}. These spectra are divided by the scaled spectral irradiance of the VUV beam.  Spectra of substrate C1 (thickness = 15 - 25 nm) show very low dispersion ($<$ 0.01 $\Delta\tau$) for consecutive measurements (C1 a) and low dispersion ($<$ 0.1 $\Delta\tau$)  between two series of measurements. Spectra of reference C2 (thickness $\sim$10 nm) show low dispersion ($<$ 0.2 $\Delta\tau$) between consecutive measurements. The films (a-C:H or soots) deposited on C1 or C2 show the expected increased optical depth with characteristic peak positions different from the substrate. Division by the respective substrate gives us the spectra in transmission for such films, as shown in Figs. \ref{fig4b} - \ref{fig:5ab}.

\end{document}